\documentclass[
twocolumn,
pre,
aps,
superscriptaddress,
showpacs,
preprintnumbers,
amsmath,
amssymb]{revtex4-1}
\usepackage{bm}

\usepackage[dvipdfmx]{graphicx}
\graphicspath{{./YAS_picture/}}
\usepackage{xcolor}

\begin{document}

\title{Identifying exogenous and endogenous activity in social media}

\author{Kazuki Fujita}
\email{fujita.kazuki.37n@st.kyoto-u.ac.jp}
\affiliation{Department of Physics, Kyoto University, Kyoto 606-8502, Japan}
\author{Alexey Medvedev}
\email{alexey.medvedev@unamur.be}
\affiliation{NaXys, Universite de Namur, 5000 Namur, Belgium}
\affiliation{ICTEAM, Universite Catholique de Louvain, 1348 Louvain-la-Neuve, Belgium}
\author{Shinsuke Koyama}
\email{koyama0526@gmail.com}
\affiliation{The Institute of Statistical Mathematics, Tokyo 190-8562, Japan}
\author{Renaud Lambiotte}
\email{renaud.lambiotte@unamur.be}
\affiliation{NaXys, Universite de Namur, 5000 Namur, Belgium}
\affiliation{Mathematical Institute, University of Oxford, Oxford, UK}
\author{Shigeru Shinomoto}
\email{shinomoto@scphys.kyoto-u.ac.jp}
\affiliation{Department of Physics, Kyoto University, Kyoto 606-8502, Japan}

\begin{abstract}
The occurrence of new events in a system is typically driven by external causes and by previous events taking place inside the system.  This is a general statement, applying to a range of situations including, more recently, to the activity of users in Online social networks (OSNs). Here we develop a method for extracting from a series of posting times the relative contributions of exogenous, e.g. news media, and endogenous, e.g. information cascade. The method is based on the fitting of a generalized linear model (GLM) equipped with a self-excitation mechanism. We test the method with synthetic data generated by a nonlinear Hawkes process, and apply it to a real time series of tweets with a given hashtag. In the empirical dataset, the estimated contributions of exogenous and endogenous volumes are close to the amounts of original tweets and retweets respectively. We conclude by discussing the possible applications of the method, for instance in online marketing. 
\end{abstract}


\maketitle

\section{Introduction}

In Online social networks (OSNs), users have the possibility to produce, consume and validate information, by posting their own content, reading the content written by others and sharing it to their own social circle \cite{bitbybit}. The growing popularity of OSNs, and the complexity and size of their data require the development of new tools for a variety of applications, going from online marketing and tracking the pulse of society \cite{Bandari2012} to sociological studies on the emergence of grassroots movement \cite{spanishprotestSciRep}. The dynamics of information in OSNs is particularly rich due to the strong heterogeneity in the users, typically associated with a broad degree distribution in the social network \cite{Kwak2010}, the competition between different keywords of hashtags \cite{Competition_memes_limitedattention,Viralityprediction_memes} and the co-existence between different types of users \cite{bot}, e.g. genuine versus bots, but also to the interplay between OSNs and more traditional mass media \cite{mass}. 

Several works have focused on the structure and dynamics of the resulting information cascades, from their characterisation in empirical data to the design of machine learning algorithms and mathematical models to predict their behaviour \cite{Lerman2010, Romero2011,zaman2014,cheng2014,Dow2013,petrovic2011rt, Aoki2016, Cattuto2007, Rybski2009}. Mathematically, information cascades are often modelled by self-exciting point processes \cite{zhao2015seismic,kobayashi2016tideh}, as previous events may trigger new events, in a way that generalises the standard Hawkes process \cite{hawkes1971}. In their simplest instance, Hawkes processes are linear self-reinforced processes, where the occurrence of an event increases the likelihood of future events. Hawkes processes have a direct connection to SI models in epidemiology \cite{Pastorsatorras2015} with, as an additional ingredient, a temporal kernel determining the stochastic time between an event and its response. This family of models has been successfully applied to model and predict, amongst others, seismic dynamics \cite{ogata1988,helmstetter2003}, scientometrics \cite{michael2012}, finance~\cite{AtSahalia2010,bacry2015, Hardiman2013} and neuronal firing~\cite{Pernice2011, ReynaudBouret2014}. 

\begin{figure*}[ht] 
\centering
\includegraphics[width=10cm]{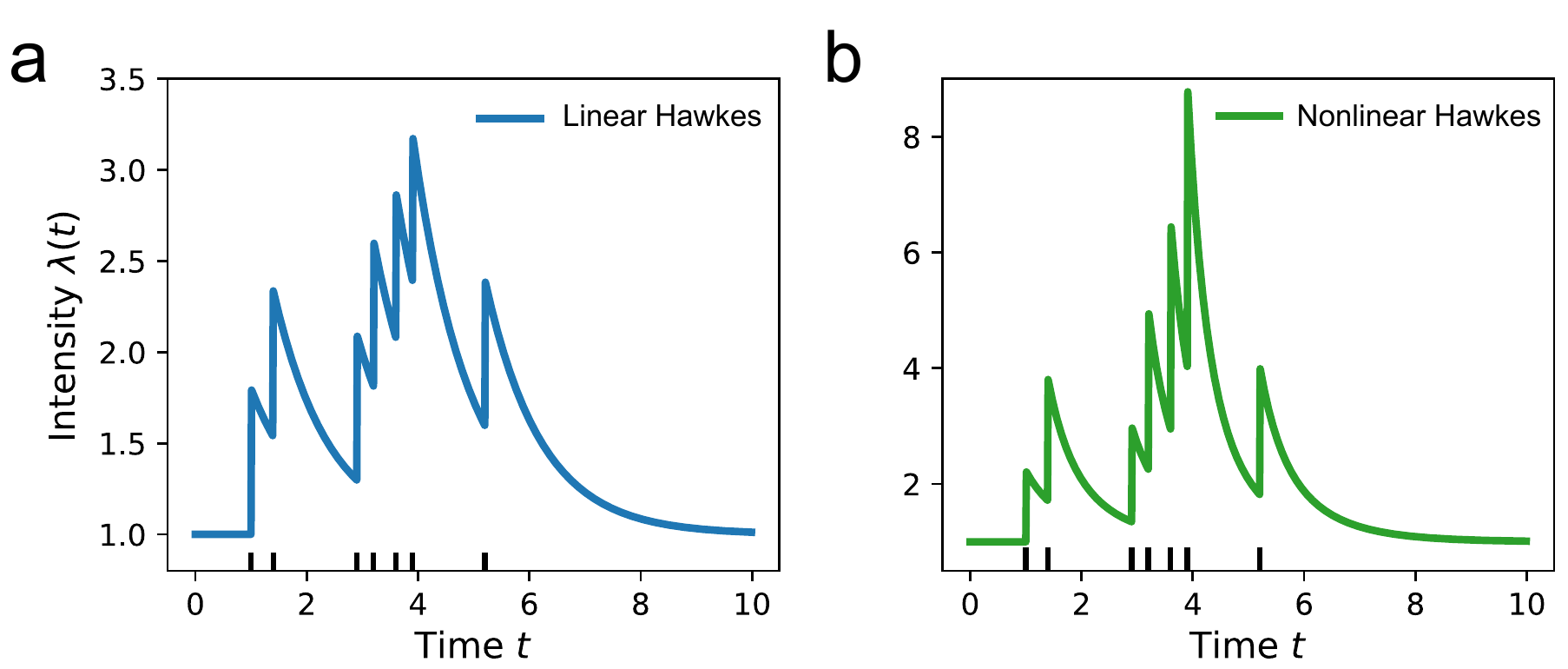} 
\caption{Comparison of intensity self-reinforcement between (a) the linear Hawkes process and (b) the non-linear Hawkes process. Both processes have background rate equals 1 and exponential memory kernel. We consider the hypothetical situation where events get realised at the same times in each case and compare the resulting intensities. Linear reinforcement generates a constant number of secondary events, while multiplicative effect is stronger if subsequent events arrived closer in time, e.g. around $t=4$.}
\label{fig:comparison}
\end{figure*}

The main purpose of this work is to design a method to identify the main forces driving the activity in an OSN, and to characterise the importance of endogenous activity, generated organically by interactions between users, and exogenous factors perturbing the internal dynamics. 
Distinguishing between exogenous and endogenous forces is critical for understanding the mechanisms that drive dynamics of OSNs and has important practical applications, such as the quantification of marketing or external factors that may manipulate the social system \cite{lazer2018science, Omi2017}. A possible solution to this challenging problem is to consider how the number of events decays after a burst of activity, as different types of relaxation are expected to emerge if the system endogenously built up its bubble of activity or if it was caused by an external shock \cite{CraneSornette2008}. However, this method suffers from practical limitations as it only allows for a post-hoc analysis after a sufficiently important burst happened. Instead of analysing gross activity, we propose to focus on the precise time series of event occurrences. Inspired by the parallels between spike train and social media time series \cite{sanli2015local}, we model the system with the generalized linear model (GLM) equipped with a self-exciting mechanism. GLMs have emerged as an important statistical framework for modelling neuronal spiking activity in a single-neuron and multi-neuronal networks \cite{Kass2018,GerhardDegerTruccolo2017} and its non-linearity presents desirable properties for information spreading on networks, as synchronised activity tends to reinforce the response to a signal. As we will show, the model naturally allows to disentangle endogenous and exogenous contributions in time series.

The rest of the paper is organized as follows. After introducing the model and the associated parameter inference, we validate the method on artificial data before testing it on empirical time series of appearance of tweets with a particular hashtag, where we successfully determine the contributions of endogenous and exogenous forces. We then provide a critical discussion about our work and conclude with possible future steps.


\section{Methods}

At the core of our method, we assume the activity time series in OSNs, for example postings of tweets with a specific keyword, is modelled by the GLM, where the underlying rate is given by
\begin{equation}
\lambda (t) = \exp \left( \gamma (t) + \alpha \sum_{k} h(t-t_k) \right),
\label{eq:nlh} 
\end{equation}
or equivalently, 
\begin{equation}
\lambda (t) = \exp \left( \gamma (t) \right) \prod_k \exp \left( \alpha h(t-t_k) \right),
\label{eq:nlh2} 
\end{equation}
where $\gamma (t)$ and $\alpha$ represent the time-varying external environment and the degree of internal self-excitation, respectively. $h(t)$ is a kernel representing the time profile of internal excitation, and $t_k$ is the occurrence time of $k$th event. Here we have chosen $h(t)=(1/\tau) \exp (-t/\tau)$ for $t>0$ and $=0$ otherwise. 

In contrast with standard linear Hawkes models, where the underlying rate has form
\begin{equation}
\lambda (t) = \mu (t) + \alpha \sum_{k} h(t-t_k),
\label{eq:lh} 
\end{equation}
the effect of previous events multiply each other, as seen in Eq.(\ref{eq:nlh2}), which results in a non-linear dynamical process. The non-linearity of the model has interesting implications for the stochastic dynamics, as it favours configurations when events appear in short bursts instead of over a long period. The model thus intrinsically rests on the importance of reinforcement, and of multiple contacts over short times to promote diffusion, as observed in complex contagion \cite{centola2010spread}, but previously modelled by means of threshold models \cite{dodds2004universal} on temporal networks \cite{takaguchi2013bursty}. This effect is illustrated in Figure \ref{fig:comparison}, where we compare examples of intensities of the linear Hawkes process and the non-linear Hawkes process of the GLM type. We observe that linear reinforcement adds a constant contribution into secondary events, while multiplicative reinforcement give a stronger push if subsequent events arrive closer in time. 

Given the occurrence rate $\lambda (t)$, the probability that events occur at times $\{t_k\} \equiv \{t_1, t_2, \cdots, t_{n} \}$ in the period of $t \in [0, T]$ is obtained as~\cite{CoxLewis1966, DaleyVereJones2003}
\begin{equation}
p \left( \{t_k\} \mid \lambda (t)\right) 
= \left[ \prod_{k=1}^{n} \lambda (t_k) \right] \exp{\left(- \int_0^T \lambda (t) \, dt \right)},
\label{eq:likelihood1} 
\end{equation}
where the exponential term is the survivor function that represents the probability that no more events have occurred in the interval.

When confronted to empirical time series, as it is usual in practice, we invert the arguments of the conditional probability Eq.(\ref{eq:likelihood1}) with Bayes' rule so that the unknown underlying rate $\lambda (t)$ is inferred from the event series observed $\{ t_k\}$:
\begin{equation}
p(\lambda (t)\mid \{t_k\}) = \frac{p(\{t_k\} \mid \lambda (t)) \,\, p(\lambda (t))}{p(\{t_k\})}.
\label{eq:posterior}
\end{equation}
As a prior distribution of $\lambda (t)$, we assume that external modulation $\gamma (t)$ is slow. This is given by penalizing the large gradient, $|d\gamma (t)/dt|$,
\begin{eqnarray}
p(\lambda (t)) \propto \exp \left[ - \beta \int_0^T \left( \frac{d\gamma (t)}{dt} \right)^2 dt \right],
\label{eq:prior}
\end{eqnarray}
where $\beta$ is a hyperparameter representing the slowness of the external fluctuations; the external stimulus is largely fluctuating if $\beta$ is small, and we interpret that external stimulus as absent if $\beta=\infty$, because $\gamma (t)$ should be constant in time in this case.

We represent as $p(\{t_k\} \mid \lambda (t))$ and $p(\lambda (t))$, respectively as $p_{\alpha}(\{t_k\} \mid \gamma (t))$ and $p_{\beta}(\gamma (t))$, by explicitly specifying the dependency on the external modulation $\gamma (t)$, internal excitation parameter $\alpha$, and the stiffness parameter $\beta$. Then the probability of having event times $p(\{t_k\})$ is given as the marginal likelihood function or the evidence:
\begin{equation}
p_{\alpha, \beta}(\{t_k\}) = \int p_{\alpha}(\{t_k\} \mid \gamma (t)) \, p_{\beta}(\gamma (t)) \, D\{\gamma (t)\},
\label{eq:marginal} 
\end{equation}
where $\int D\{\gamma (t)\}$ represents a functional integration over all possible paths of external fluctuations $\gamma (t)$. The method of selecting the hyperparameters according to the principle of maximizing the marginal likelihood function is called the Empirical Bayes method~\cite{Good1966, Akaike1980, MacKay1992, CarlinLouis2000}. The marginalization path integral Eq.(\ref{eq:marginal}) for a given set of time series $\{t_k\}$ can be carried out by the Expectation Maximization (EM) method~\cite{DempsterLaird1977, SmithBrown2003} or the Laplace approximation~\cite{KoyamaPaninski2010}.

In this framework, the contributions of endogenous and exogenous origins that have influenced for the occurrence of events are judged by the hyperparameters $\{\hat{\alpha}, \hat{\beta}\}$ selected by maximizing the marginal likelihood $p_{\alpha, \beta}(\{t_k\})$ (Table {\ref{table1}}). Given the hyperparameters determined as $\{\hat{\alpha}, \hat{\beta}\}$, we can obtain the maximum \textit{a posteriori} (MAP) estimate of the external circumstance $\hat{\gamma}(t)$, with which their posterior distribution, 
\begin{equation}
p_{\hat{\alpha}, \hat{\beta}}(\gamma (t)\mid \{t_k\}) \propto p_{\hat{\alpha}}(\{t_k\} \mid \gamma (t)) \, p_{\hat{\beta}}(\gamma (t)),
\label{eq:map}
\end{equation}
is maximized. With the estimated $\hat{\gamma}(t)$ and the given series of event times $\{ t_k\}$, we obtain the rate $\lambda_{\rm GLM} (t)$ as 
\begin{equation}
\lambda_{\rm GLM} (t) = \exp \left( \hat{\gamma} (t) + \hat{\alpha} \sum_{k} h(t-t_k) \right).
\label{eq:glmrate} 
\end{equation}

\begin{table} []
\caption{Presence of endogenous and exogenous contributions can be inferred by the selected hyperparameters of the GLM.}
\begin{center}
\begin{tabular}{| c || c | c| }
\hline
& \,\, self-excitation $\hat{\alpha}$ \,\, & \,\, stiffness $\hat{\beta}$ \,\, \\ \hline 
\,\, endogenous \,\, & \,\, finite\,\, & $ \,\, \infty \,\, $ \\ \hline
\,\, exogenous \,\, & \,\, 0 \,\, & \,\, finite \,\, \\ \hline
\,\, endo. + exo. \,\, & \,\, finite \,\, & \,\, finite \,\, \\
\hline
\end{tabular}
\end{center}
\label{table1}
\end{table}

\begin{figure*}[htb] 
\centering
\includegraphics[width=11.5cm]{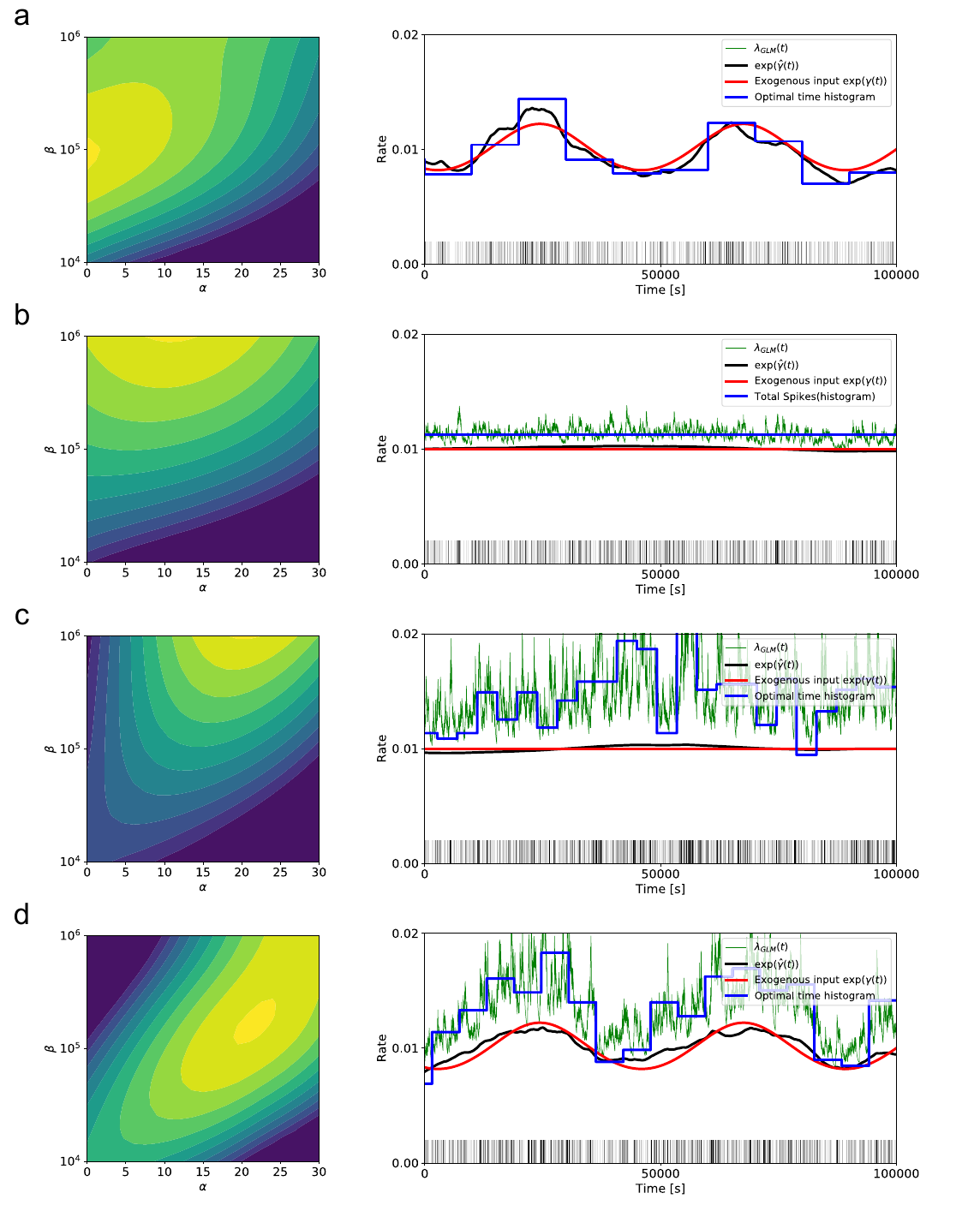} 
\caption{ Fitting GLM to synthetic data. (a) Exogenously modulated rate process (Eq.(\ref{eq:extrinsic}), $\gamma_0=\log{0.01}$; $b_0=0.2$; $T=43200$). (b) The nonlinear Hawkes process with small self-excitation (Eq.(\ref{eq:intrinsic}), $\gamma_0=\log{0.01}$; $\alpha_0=10$; $\tau_0=300$). (c) The nonlinear Hawkes process with the larger self-excitation (Eq.(\ref{eq:intrinsic}), $\gamma_0=\log{0.01}$; $\alpha_0=25$; $\tau_0=300$). (d) The system receiving external fluctuations and self-excitation (Eq.(\ref{eq:intextrinsic}), $\gamma_0=\log{0.01}$; $b_0=0.2$; $T=43200$; $\alpha_0=20$; $\tau_0=300$). (left panel) Contour plot of the log-likelihood (Eq.(\ref{eq:marginal})). (Right panel) The solid blue line shows the optimal time histogram, red line below shows the original exogenous activity $\exp(\gamma (t))$, black curve represents the inferred exogenous activity $\exp(\hat{\gamma} (t))$, and green is the total rate given by the GLM $\lambda_{\rm GLM} (t)$ (Eq.(\ref{eq:glmrate})).}
\label{fig:synthetic}
\end{figure*}

\section{Results}

\subsection{Application to synthetic data}

Here we test the efficiency of the method by fitting it to series of occurrence times derived from the following rate processes: 
\begin{enumerate}
\item[(a)] {\bf [exogenous modulation]} Firstly we considered an inhomogeneous Poisson process in which events are drawn from a time varying rate:
\begin{equation}
\lambda (t) = \exp \left(\gamma_0 + b_0 \sin(t/T)\right).
\label{eq:extrinsic}
\end{equation}
We interpret this mode as purely exogenous because the rate variation is independent of past events. We fit our GLM to a series of occurrence times derived from this rate process. The left panel of Figure~\ref{fig:synthetic}(a) shows a contour plot of the log-likelihood (Eq.(\ref{eq:marginal})), indicating that the self-excitation parameter $\hat{\alpha}$ is zero while the stiffness constant $\hat{\beta}$ is finite. Thus the method suggests that the rate modulation would have been exogenous. In the right panel of Figure~\ref{fig:synthetic}(a), the occurrence rate estimated with our GLM, $\exp (\hat{\gamma}(t))$, is compared with a time histogram optimally fitted to the data~\cite{ShimazakiShinomoto2007}, demonstrating that the GLM has succeeded in capturing the underlying rate properly.

\begin{figure*}[]
\centering
\begin{minipage}[b]{0.3\linewidth}
\centering
\includegraphics[width=0.8\linewidth]{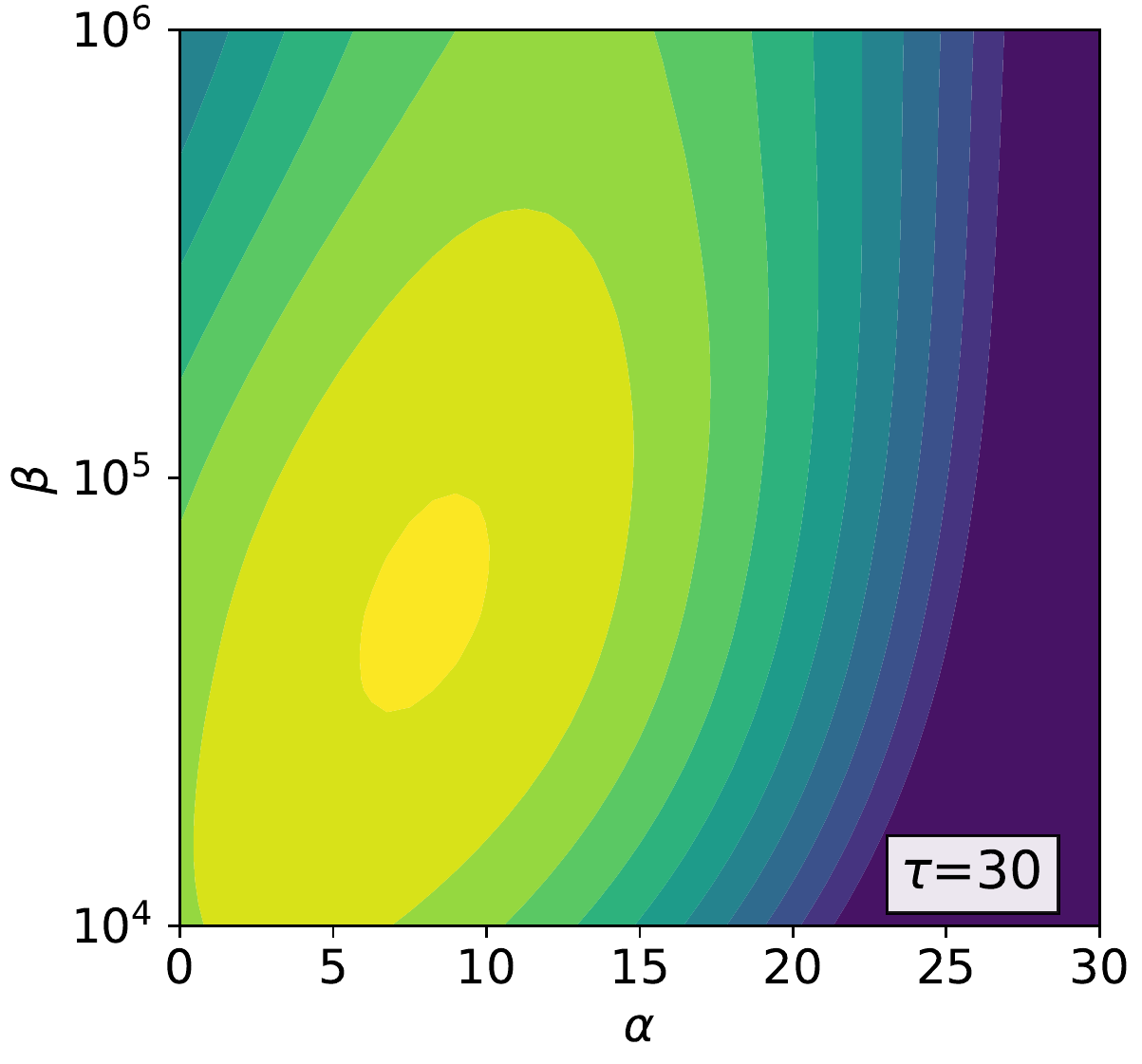}
\end{minipage}
\begin{minipage}[b]{0.3\linewidth}
\centering
\includegraphics[width=0.8\linewidth]{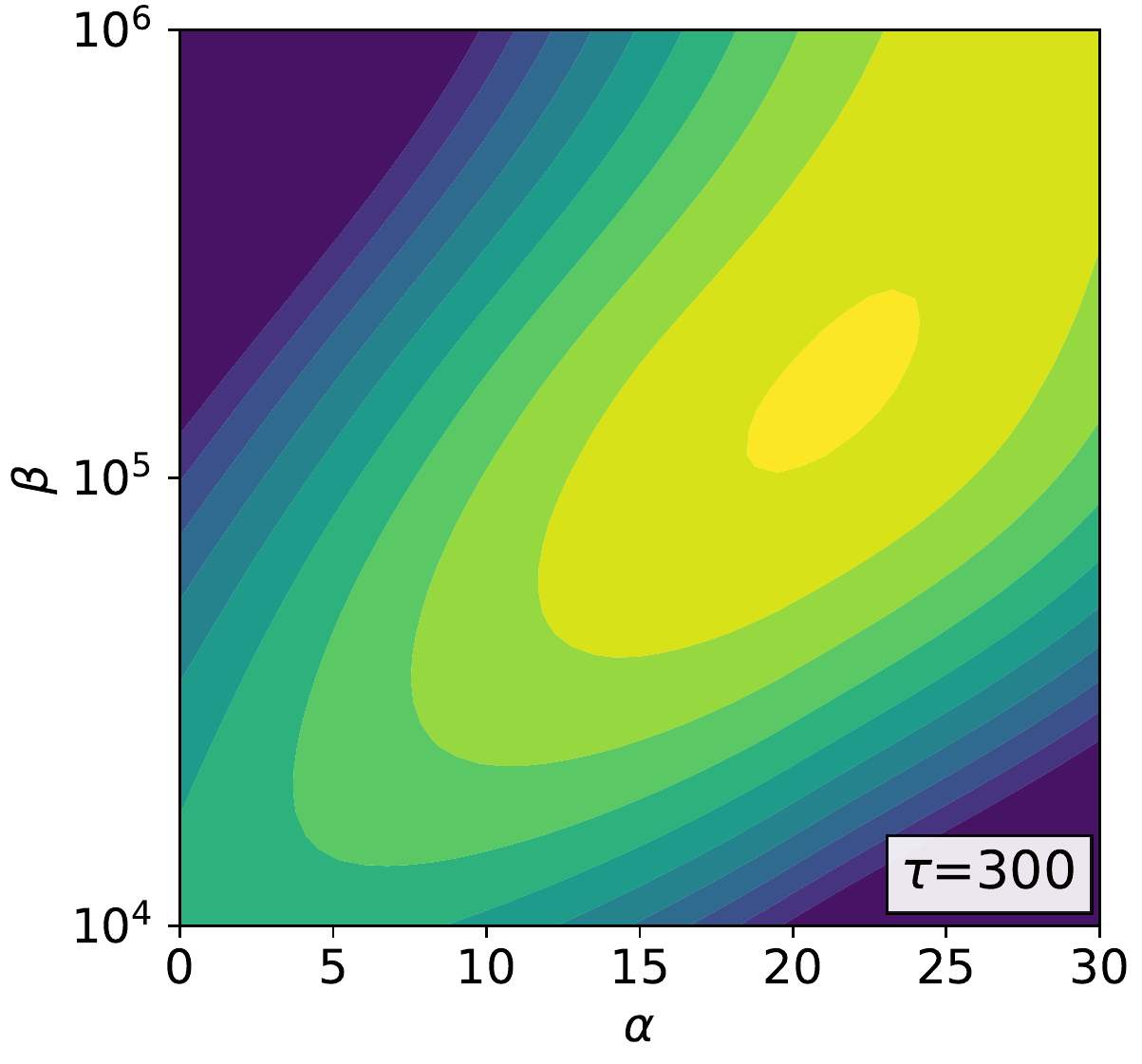}
\end{minipage}
\begin{minipage}[b]{0.3\linewidth}
\centering
\includegraphics[width=0.8\linewidth]{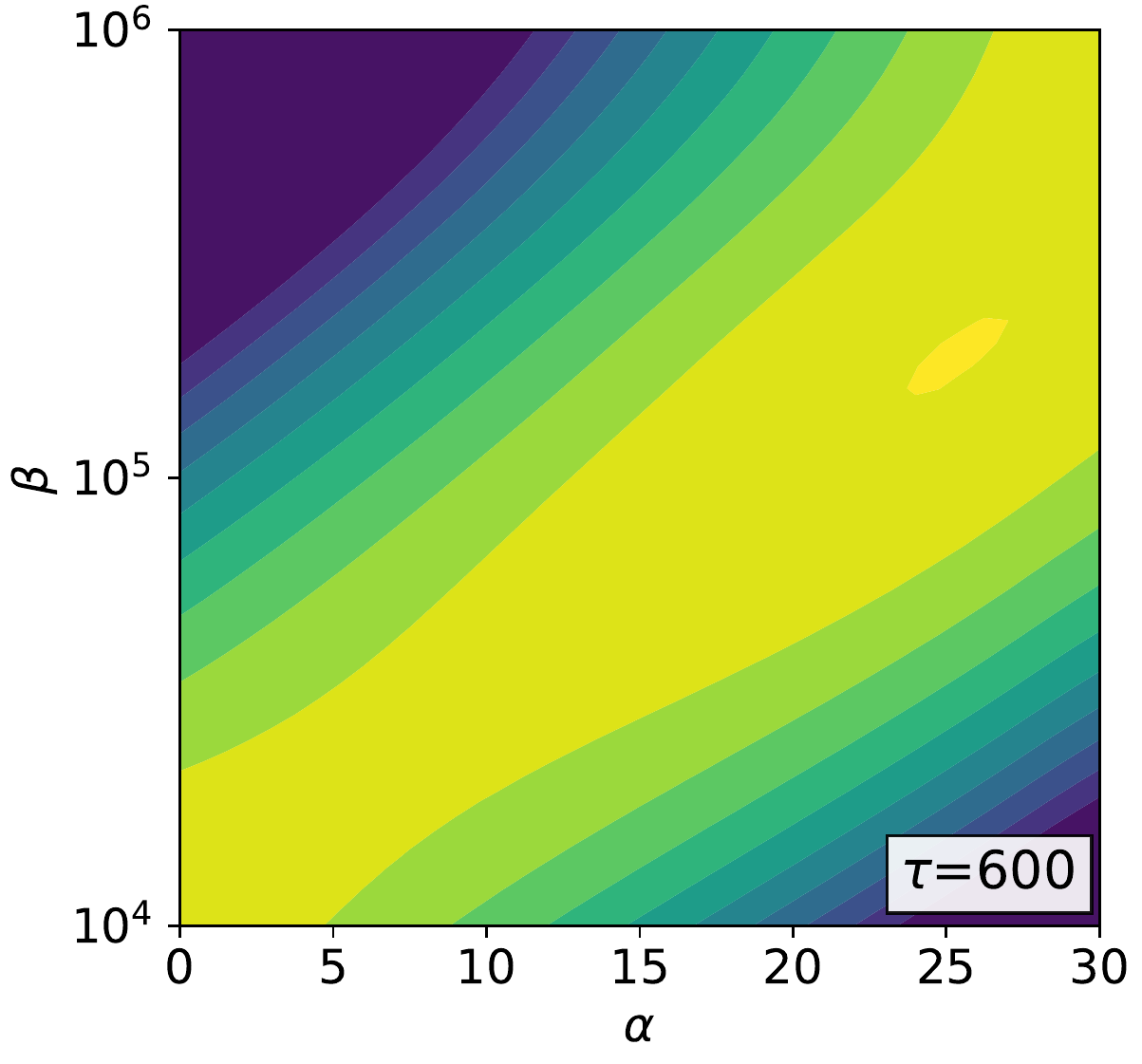}
\end{minipage}
\caption{Contour plots of the log-likelihood function obtained with the GLM timescales $\tau=30$, $300$, and $600$ seconds. The original data was derived from the nonlinear Hawkes process of the system receiving external fluctuations and self-excitation of the timescale $\tau_0=300$ seconds as in Figure~\ref{fig:synthetic}(d).} 
\label{fig:contour_dt}
\end{figure*}

\item[(b)] {\bf [endogenous modulation with a small self-excitation]} We generated events with the nonlinear Hawkes process
\begin{equation}
\lambda (t) = \exp \left(\gamma_0 + \alpha_0 \sum_{k} h_0(t-t_k) \right),
\label{eq:intrinsic}
\end{equation}
where we have taken the kernel $h_0(t)=(1/\tau_0) \exp (- t/\tau_0)$ for $t>0$ and $=0$ otherwise. Here, we have chosen the timescale of the GLM kernel ($\tau$) as identical to the timescale of this generative model ($\tau_0$). By applying our GLM to the series of occurrence times, the self-excitation parameter $\hat{\alpha}$ is selected as non-zero, suggesting that the system had endogenous excitation (Figure~\ref{fig:synthetic}(b) left panel). Because the stiffness $\hat{\beta}$ is very large, the base rate $\exp (\hat{\gamma}(t))$ is nearly constant, indicating that external circumstances were stationary. The total rate estimated by the GLM, Eq.(\ref{eq:glmrate}), is very close to the original rate given in Eq.(\ref{eq:intrinsic}). For this data, the optimized bin size of the histogram diverges, indicating that the fluctuation in the rate was not detected. The estimated (constant) rate is above the baseline rate $\exp(\hat{\gamma})$, because the contribution of the self-excitation is included in the total rate (Figure~\ref{fig:synthetic}(b) right panel). 

\item[(c)] {\bf [endogenous modulation with a larger self-excitation]} We generated events with the nonlinear Hawkes process given by Eq.(\ref{eq:intrinsic}) with the self-excitation term $\alpha_0$ greater than the case (b), so that event occurrence exhibits large fluctuations. By applying the optimal histogram method, we obtained fluctuating rate (i.e., the optimal bin size was finite), implying that the nonlinear Hawkes process may also exhibit the stationary-nonstationary (SN) transition, which was found in the linear Hawkes process~\cite{OnagaShinomoto2014, OnagaShinomoto2016}: significant fluctuations appear even in the absence of external modulation. Although the rate estimation method suggested that rate is fluctuating, our GLM was able to see through that exogenous forcing was absent, and conclude that the fluctuations appeared solely due to the self-excitation (Figure~\ref{fig:synthetic}(c) right panel). 

\item[(d)] {\bf [exogenous + endogenous modulation]} We derived events from the system receiving both external fluctuations and self-excitation:
\begin{equation}
\lambda (t) = \exp \left(\gamma_0 + b_0 \sin(t/T)+\alpha_0 \sum_{k} h_0(t-t_k) \right).
\label{eq:intextrinsic}
\end{equation}
By applying our GLM to a series of occurrence times, we obtain the self-excitation parameter $\hat{\alpha}$ and stiffness constant $\hat{\beta}$ both finite, as shown in the contour plot of the log-likelihood (Figure~\ref{fig:synthetic}(d) left panel), suggesting that the system would have been stimulated exogenously but there would have been the endogenous self-excitation mechanisms either. 
\end{enumerate}

In the above, we have seen that the GLM is able to decipher the original self-excitation mechanisms provided that the event generation process (the nonlinear Hawkes process in this case) is contained in a family of rate processes presumed for the GLM. In real applications, however, the precise underlying mechanisms of data generation are usually hidden, thus accordingly we have to assume that our GLM may not cover the original process. To examine whether or not the GLM may work even if the model does not contain the original process, we performed the following tests: We generated a series of events from a nonlinear Hawkes process with exponential self-excitation kernel and timescale $\tau_0=300$ seconds (Eq.(\ref{eq:intextrinsic}), $(\alpha, \beta)$ both finite), and fitted GLMs whose self-excitation timescale $\tau$ is different from $\tau_0$. We confirmed that the GLM suggests finite optimal $(\hat{\alpha},\hat{\beta})$ for a rather wide range of timescales $\tau$ (between 10 and 600 seconds). Figure~\ref{fig:contour_dt} displays contour plots of the log-likelihood function. This implies that the GLM may capture the presence of self-excitation and external fluctuation robustly even if the precise temporal profile of the self-excitation is not \textit{a priori} known.


\begin{figure*}[]
\includegraphics[width=15cm]{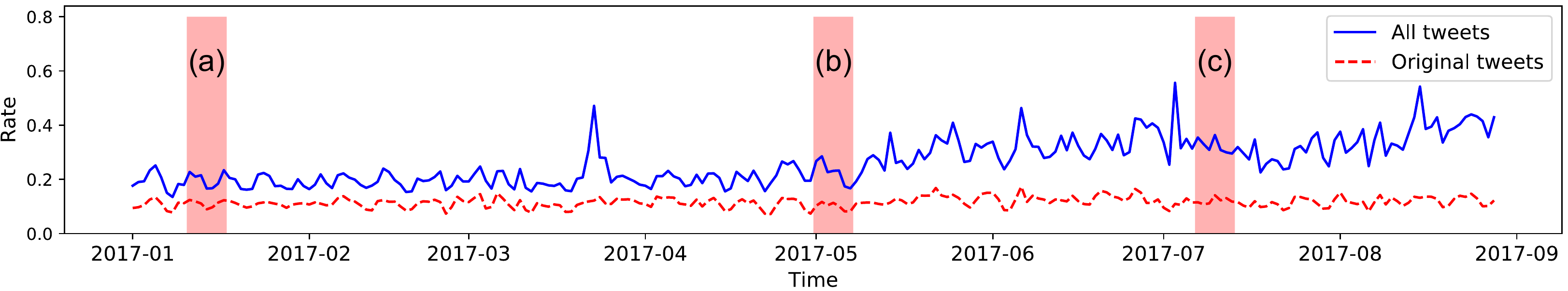}
\caption{Tweeting rate in the whole dataset from Jan till Sept 2017. The solid blue line shows tweeting rate of the tweets that contain a hashtag, related to `bitcoin'. Dashed red line shows the rate of appearance of original tweets with the same hashtags. Both rates were approximated from daily bins. }
\label{fig:whole_dataset}
\end{figure*}

\begin{figure*}[htb]
\includegraphics[width=15cm]{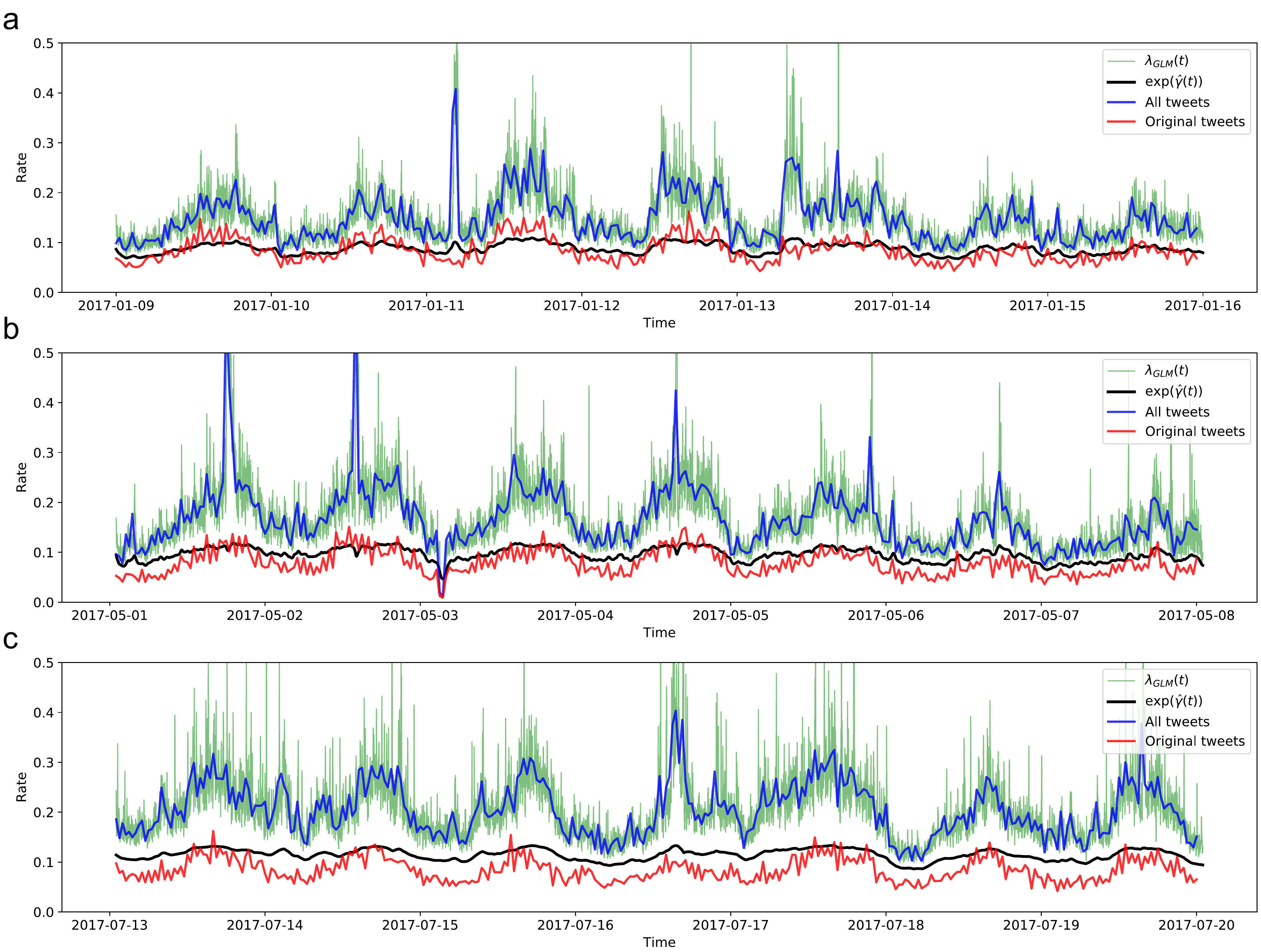}
\caption{Original tweeting rate estimation for the one week samples of tweets: (a) between Jan 9 and Jan 16, 2017, (b) between May 1 and May 8, 2017, and (c) July 13 - July 20, 2017. The solid blue line shows the tweeting rate of all tweets, red line below shows the rate of original tweets, black curve represents the inferred exogenous activity $\exp(\hat{\gamma} (t))$ and green is the total rate given by the GLM. The total and original tweeting rates were approximated using 20 min bins and the estimated rates are given using 1 min bins.}
\label{fig:result_full}
\end{figure*}

\begin{figure*}[]
\centering
\begin{minipage}[b]{0.3\linewidth}
\centering
\includegraphics[width=0.8\linewidth]{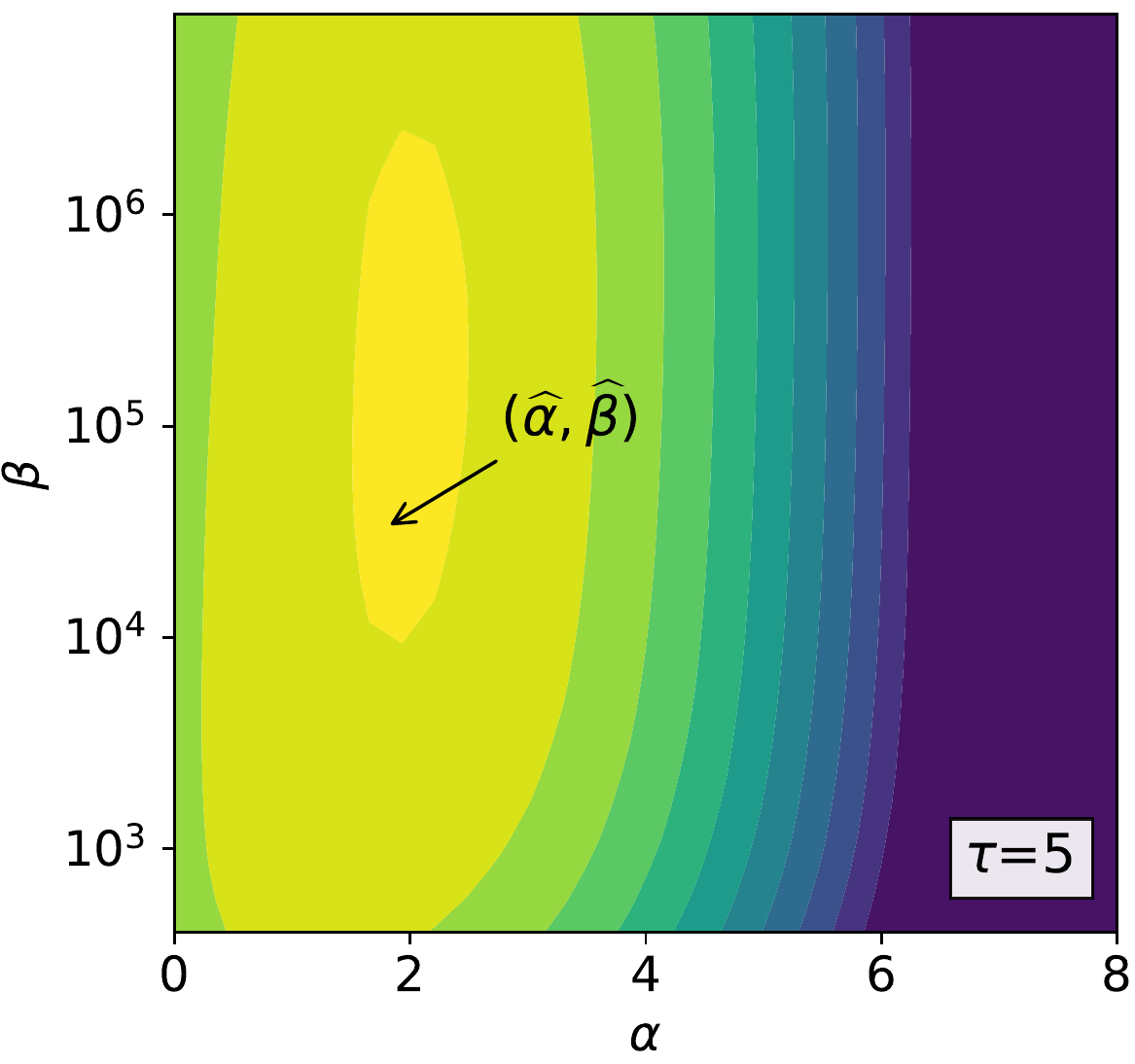}
\end{minipage}
\begin{minipage}[b]{0.3\linewidth}
\centering
\includegraphics[width=0.8\linewidth]{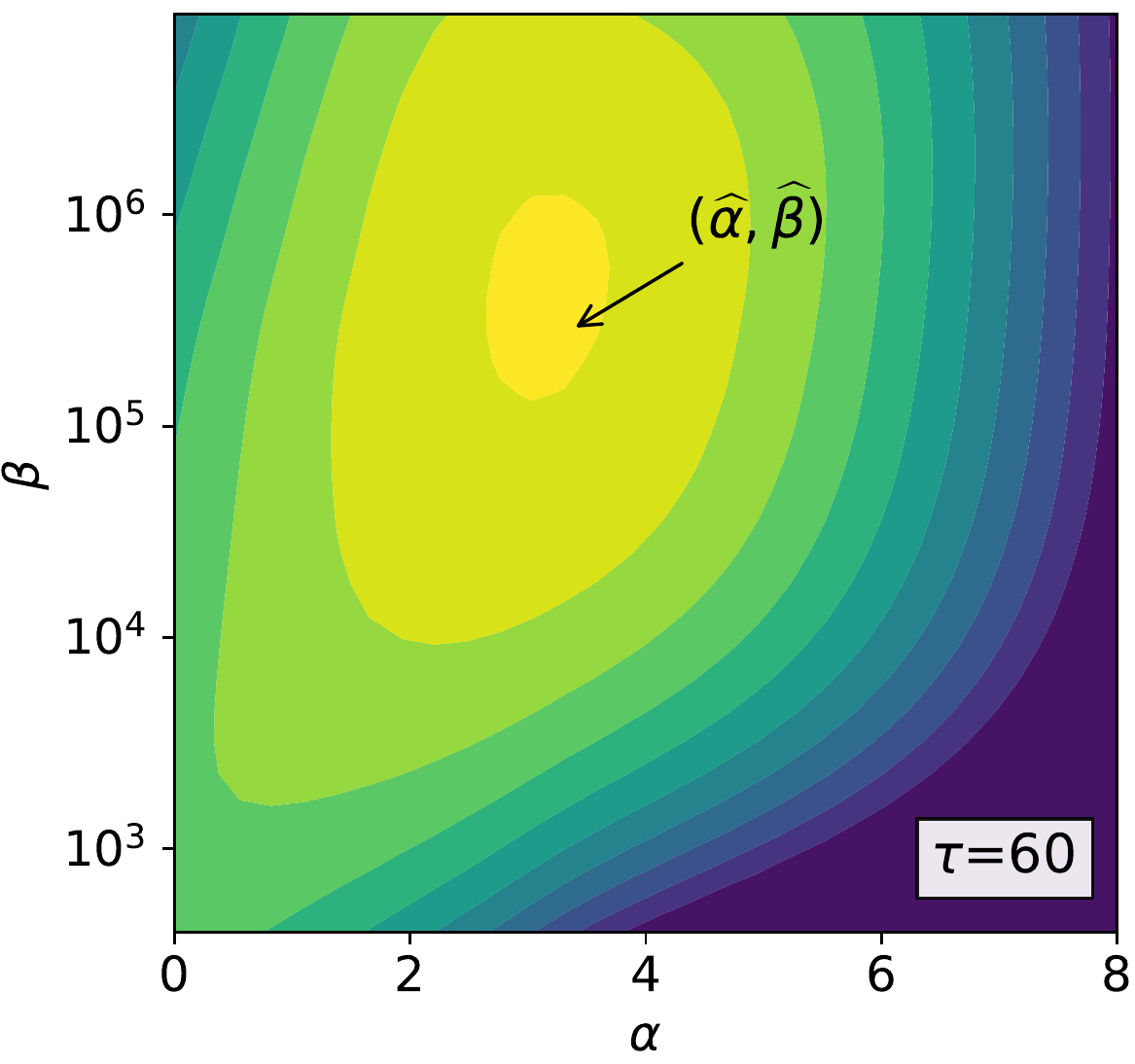}
\end{minipage}
\begin{minipage}[b]{0.3\linewidth}
\centering
\includegraphics[width=0.8\linewidth]{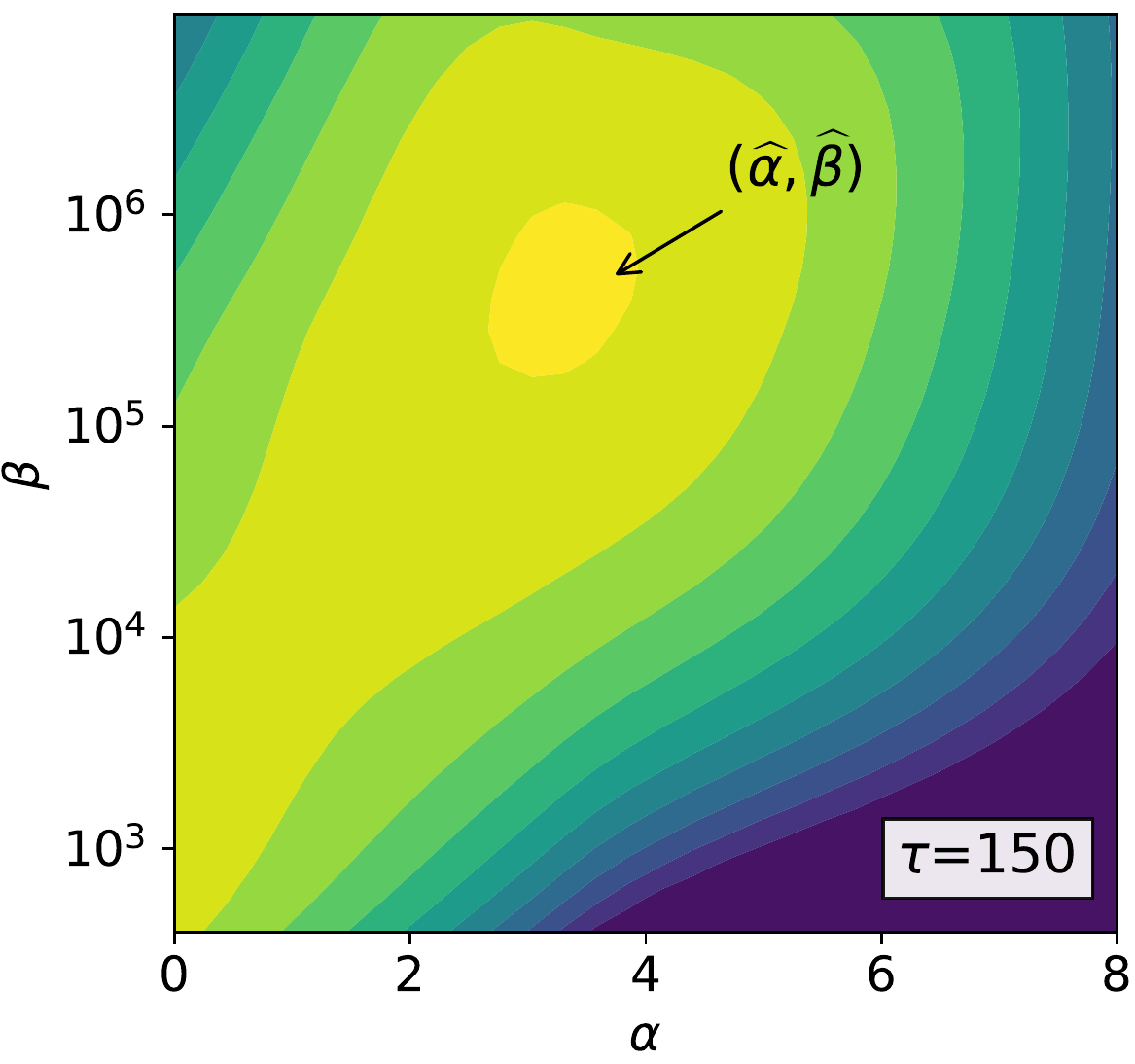}
\end{minipage}
\caption{Contour plots of the log-likelihood function for timescales $\tau=5,60$, and $150$ seconds. The optimal value found by GLM method is depicted as $(\widehat{\alpha},\widehat{\beta})$. }
\label{fig:contour_tweet}
\end{figure*}

\subsection{Application to a series of tweet times}

In an OSN, one may differentiate between the production of original content and the sharing of existing content over the network of peers. Content may be related to a topic or a real-world event, and its appearance  in the digital space is modulated by its interest. When considering the total number of occurrences of a topic-related content, one may interpret the original posts as exogenous input, since the content arrives extrinsically into the social system, while following reshares or retweets  may be considered as an endogenous self-exciting contribution into dynamics. These two processes are undoubtedly coupled together, thus it is hard to directly separate one type of activity from the other by observing only the global time series. 

We test our separation method on the data from Twitter, which is a perfect example of content sharing social system. We consider the dataset of tweets, collected through the public API, posted between January and late August of 2017 that contain the hashtag \#bitcoin. These tweets represent the topic of one cryptocurrency and public attention to it. The dataset contains 13,365,114 tweets and for each tweet we have information about its creation time, its content and whether the tweet is an original piece of content or a retweet. Note that no underlying network of followers was captured. From this information we infer two separate time series, one related to the \textit{original} tweet postings with the hashtag and another represents the \textit{total} hashtag appearance, including retweets. The average rate of appearance of these two types of tweets is drawn on Figure~\ref{fig:whole_dataset}. Both rates were approximated from daily bins for the sake of clarity of presentation. We observe an increase in appearance rate of retweeted content while the rate of original tweets remains practically stable. Since the tweets are related to the topic of cryptocurrencies, this may be explained by a growing attention to bitcoin related to its recent growth in volume and market capitalisation~\cite{CoinMarketCap2018}. 

We apply our GLM in order to separate the original tweeting rate from retweeting. Due to the large size of the observation window we select three one week samples from the dataset and present our analysis on these samples (Figure~\ref{fig:whole_dataset}). We applied the model to other samples from the dataset and the results were comparable and are not shown here due to space limitation. Following the rapid nature of retweeting activity~\cite{zhao2015seismic, kobayashi2016tideh}, we use the exponential kernel with timescale parameter $\tau$ \textit{a priori} set to $60$ seconds. As to the data examined, time stamps were recorded in seconds and data contains a non-zero fraction of multiple timestamps falling into the same second. We confirm that randomization of these multiple events in a half-second radius around the given second timestamp performed worse than simply disregarding them. Therefore, in our experiment we stick to the latter option.

The results of the exogenous activity separation are shown on Figures~\ref{fig:result_full}. For the sake of clarity of presentation, the original and total tweeting rates were shown using the 20 min binning of time stamps and estimated tweeting rates are shown using 1 min binning. We first observe that large peaks in the total tweeting activity are not accompanied by peaks in the rate of original tweets arrival, therefore those are clearly due to retweets. The GLM method succeeded in filtering out these bursts of activity and the estimated exogenous rate $\exp(\hat{\gamma} (t))$ is close to the rate of original tweets. The total estimated rate $\lambda_{\rm GLM}(t)$ shows to precisely follow the total tweeting activity, which is though expected, since the algorithm optimizes the difference between total tweeting rate and $\lambda_{\rm GLM}(t)$. However, there appears to be a slight discrepancy in the Figure~\ref{fig:result_full}, (c), which may be explained by the growth of attention in combination with one second resolution drawback. The contour plots for the time series (a) show clear finite optimal $(\widehat{\alpha}, \widehat{\beta})$ for various values of timescale parameter $\tau$ (Figure~\ref{fig:contour_tweet}). 

\section{Discussion}

We have developed the GLM-based method to estimate the influence of exogenous and endogenous forces on observed temporal events. Using synthetic data generated by non-linear Hawkes processes, we confirmed that the method is capable of estimating the respective contributions. Then we applied the method to the time series of tweets with a given hashtag, and found that the estimated contributions of external and internal origins are close to the original tweets and retweets, respectively. 

The concept of dividing the world into exogenous and endogenous categories is a controversial philosophical problem, and it might be considered as a subjective decision. However, the estimation of the exogenous component from a time series has important implications to design efficient models to predict the future of a time series and to infer the impact of a marketing campaign on the activity of a social network, judging whether items require extensive advertisement or word-of-mouth product mentions have already gone viral.

Note that another method has been designed for a similar purpose, based on the fitting of the linear Hawkes process using the EM method, and validated on a data set of violent civilian deaths occurring in the Iraqi conflict~\cite{LewisMohler2012}. An advantage of this method is the linearity of the model, which avoids possible catastrophic divergences in the number of events~\cite{GerhardDegerTruccolo2017}. However, our GLM-based approach has the advantage of determining the timescale of exogenous fluctuation $\beta$ semi-automatically, according to the Empirical Bayes method, while this timescale needs to be given manually in the linear model. For these reasons, our method is expected to perform well in situations when the exogenous activity has a slow modulation. Because there are many cases in which external stimuli are given, abruptly triggering the following responses, it is worthwhile to develop a method of analysing such cases. 

The continuous nature of the GLM suggests the recorded data is continuous as well. However, in practice the high precision temporal data is rarely available, usually the time is rounded up to a second. Thus the drawback of multiple events may occur in case when the collected time series come from a process of high frequency, which can be subdued by improving the data measurement frequency. On the other hand, it would increase computational time of the algorithm, a classic precision/speed trade-off. Another practical issue is selection of the self-excitation kernel and its timescale $\tau$. The proposed method showed to succeed when the provided value of $\tau_0$ lies in a certain interval around the true $\tau$ of the process. Narrowing this interval down to a correct $\tau$ can be done using extra available information, e.g. retweet time distribution of a test sample of tweets. 

\vspace{5mm}

\acknowledgments

This study was supported in part by Grants-in-Aid for Scientific Research to SS from JSPS KAKENHI Grant numbers 26280007 and 17H06028 and JST CREST Grant Number JPMJCR1304. AM, RL and SS were supported by the Bilateral Joint Research Project between JSPS, Japan, and FRS-FNRS, Belgium. AM was supported by ARC (Federation Wallonia-Brussels) and by the Russian Foundation of Basic Research 16-01-00499.


\bibliography{references}
\end{document}